# Multipacting mitigation by atomic layer deposition: the case study of Titanium Nitride.

Y. Kalboussi[1], S. Dadouch[2], B. Delatte[1], F. Miserques[3], D. Dragoe[4], F. Eozenou[1], M. Baudrier[1], S. Tusseau-Nenez[5], Y. Zheng[6], L. Maurice[1], E. Cenni[1], Q. Bertrand[1], P. Sahuquet[1], E. Fayette[1], G. Jullien[1], C. Inguimbert[2], M. Belhaj[2], and T. Proslier[1].

[1] Département des Accélérateurs, de la Cryogénie et du Magnétisme, Université Paris-Saclay, CEA, 91191 Gif-sur-Yvette, France.
[2] Département Physique, instrumentation, environnement, espace, ONERA, 31055 Toulouse, France
[3] Service de Recherche sur la Corrosion et le Comportement des Matériaux, Université Paris-Saclay, 91191 Gif sur Yvette, France.
[4] Institut de Chimie Moléculaire et des Matériaux d'Orsay, Université Paris-Saclay, 91400 Orsay, France.
[5] Laboratoire de Physique de la Matière Condensée, UMR 7643 CNRS-Ecole polytechnique, France
[6] Institut des NanoSciences de Paris, UMR7588, CNRS- Sorbonne Université, France.

**Abstract**

This study investigates the use of Atomic Layer deposition (ALD) to mitigate multipacting phenomena inside superconducting radio frequency (SRF) cavities used in particle accelerators. The unique ALD capability to control the film thickness down to the atomic level on arbitrary complex shape objects enable the fine tuning of TiN film resistivity and total electron emission yield (TEEY) from coupons to devices. This level of control allows us to adequately choose a TiN film thickness that provide both a high resistivity to prevent Ohmic losses and low TEEY to mitigate multipacting for the application of interest. The methodology presented in this work can be scaled to other domain and devices subject to RF fields in vacuum and sensitive to multipacting or electron discharge processes with their own requirements in resistivities and TEEY values.

## 1. Introduction

The total electron emission yield (TEEY) is a well-known phenomenon of electron-solid interaction where primary incident electrons hitting a surface, induce the emission of electrons: secondary (SEE) and backscattered (BSE) electrons. SEE and BSE has been considered as accounting for the parasitic multipactor effect in Microwave systems. Multipactor is a resonant electron discharge caused by the synchronization of the emitted electrons with the electric field and their uncontrolled multiplication at each impact with the surface.

The physical mechanism behind this effect is an avalanche caused by the electrons emitted: A primary electron impacts the surface and depending on its energy and total electron emission yield (TEEY) of the material, a number of secondary and backscattered electrons will be released from the surface, which may in terms be accelerated by the reversed radio frequency (RF) fields and impact the surface releasing even more electrons and so on. This detrimental effect takes place in an extremely wide range of devices extending from divertors in tokamaks (Gunn, 2012), to space satellites (A. V. Streltsov, 2018), antennas (Golio, 2001), power couplers (Jerzy Lorkiewicz, 2004) and radio-frequency cavities (SRF) (R. Prakash, 2017) (J Knobloch, 1997) in particle accelerators and can cause consequent energy deposition, leading for instance to SRF cavity quenches (Proch, 1979), power coupler window breakage, and in severe cases, vacuum breakdown and the destruction of the RF device.

Consequently, intense research efforts, in all these fields, are pursued to reduce the TEEY of the surface. Coating the surface with intrinsically low TEEY films such as titanium nitride (A. Variola, 2008), chromium oxides (J. Lorkiewicz, 2004) and carbon (M. Angelucci, 2020) were found to be one of the most effective routes to suppress multipacting.

Coating an RF surface to decrease its TEEY may however meet with other necessarily design specifications: the thin film resistivity in particular can severely affect the RF surface losses. For instance, state of the art superconducting Radio Frequency (SRF) accelerating cavities used in particle accelerators have reproducible surface resistances of few nano Ohms or quality factors (Q) above $10^{10}$ (J Knobloch, 1997) and the presence of non-superconducting, metallic layers within the RF skin depth will increase the surface resistance and decrease the Q drastically. Hence in order to mitigate the multipacting effects in SRF cavities the coating properties requirements are both a low TEEY and transparent to RF, i.e insulating. Other applications such as ceramic windows in an intense RF field and vacuum environments (power couplers, RF wave guides) require on the contrary the charge evacuation to prevent charging and arc formations and a low TEEY. The charge evacuation is achieved by finely tuning the thin film resistivity that can be achieved by controlling its thickness and chemical composition.

Another important aspect to take into consideration is the coating uniformity over potentially complex shaped devices surfaces, such as particle accelerator accelerating cavities, antennas or 3D printed embarked structures.

For these particular reasons, we use thermal Atomic layer deposition (ALD) as a deposition technique widely known for its excellent uniformity and atomic-level thickness control (R.W. Johnson, 2014) (M. Leskelä, 2003). ALD is a chemical gas phase synthesis method used in microelectronics (C. Wiemer, 2012), photovoltaics (J.A. Van Delft, 2012) and battery (X. Meng, 2012) based on cyclic, self-saturating gas surface chemical reactions. It is important to highlight that, unlike other deposition techniques such as evaporation and sputtering where molecules or nanoparticles agglomerate randomly on the surface until they form a continuous film, during an ALD process the precursors molecules reacts covalently with the substrate through a self-limiting chemical reaction and leaves no more than one monolayer precursor molecules at the surface after each ALD cycle (George, 2010). The film is therefore deposited one monolayer at a time and this stages for an outstanding thickness and chemical composition control and uniformity over other deposition techniques such as chemical vapor deposition (CVD), physical vapor deposition (PVD) and sputtering.

In this paper, we investigate the thickness dependence of the total electron emission yield (TEEY), the chemical composition and resistivity of titanium nitride nano layers deposited by thermal ALD and their response to electron bombardment conditioning. The main findings of this study show that for a TiN film thicknesses above 1.5-2 nm, the maximum TEEY reach a saturating "bulk" value of 1.85 and 1.25 before and after conditioning. For comparison, other deposition techniques show a saturation of the max TEEY above few tens of nanometers on rough ceramic surfaces and highlights the unique film thickness control and uniformity capabilities of ALD as previously measured on $Al_2O_3$ and MgO films with different thicknesses (S.J. Jokela, 2012). In addition, the sub-nanometer control of the TiN film thickness enable a fine tuning of both the electrical resistivity and the TEEY values of the TiN films. As a proof of concept, the growth process parameters were optimized on coupons to obtain targeted TEEY and resistivity values for SRF cavities application. These parameters were then successfully implemented on a 1.3 GHz cavity used in particle accelerators in our custom built ALD system. The RF tests conducted at 1.5 K revealed that 1.5 nm of TiN efficiently suppress multipacting while preserving high quality factors in the $10^{10}$.

## 2. Material and methods

### 2.1. Atomic layer deposition method

The TiN films were deposited on a 8 nm thick film of ALD $Al_2O_3$. Both $Al_2O_3$ and TiN films were deposited in situ by thermal ALD in a home built viscous-flow thermal ALD reactor (J.W. Elam, 2002) with a 5 cm diameter and 50 cm long deposition chamber. The reactor temperature was maintained during the deposition process by a computer-monitored resistive heater system in a range of 30°C to 500°C. Several K-type thermocouples were placed along the length of the flow tubes and the deposition chamber to ensure the temperature uniformity. The carrier gas used was ultrahigh-purity nitrogen (UHP, 99.999 %) further purifier using an inert gas purifier. The flow was maintained constant during the deposition at 250 sccm with a reactor pressure of 0.9 ± 0.05 mbar. The ALD system is also equipped with a residual gas analyzer (RGA) device to monitor the reactor gas chemical composition and hence following the surface chemicals reactions inside the deposition chamber.

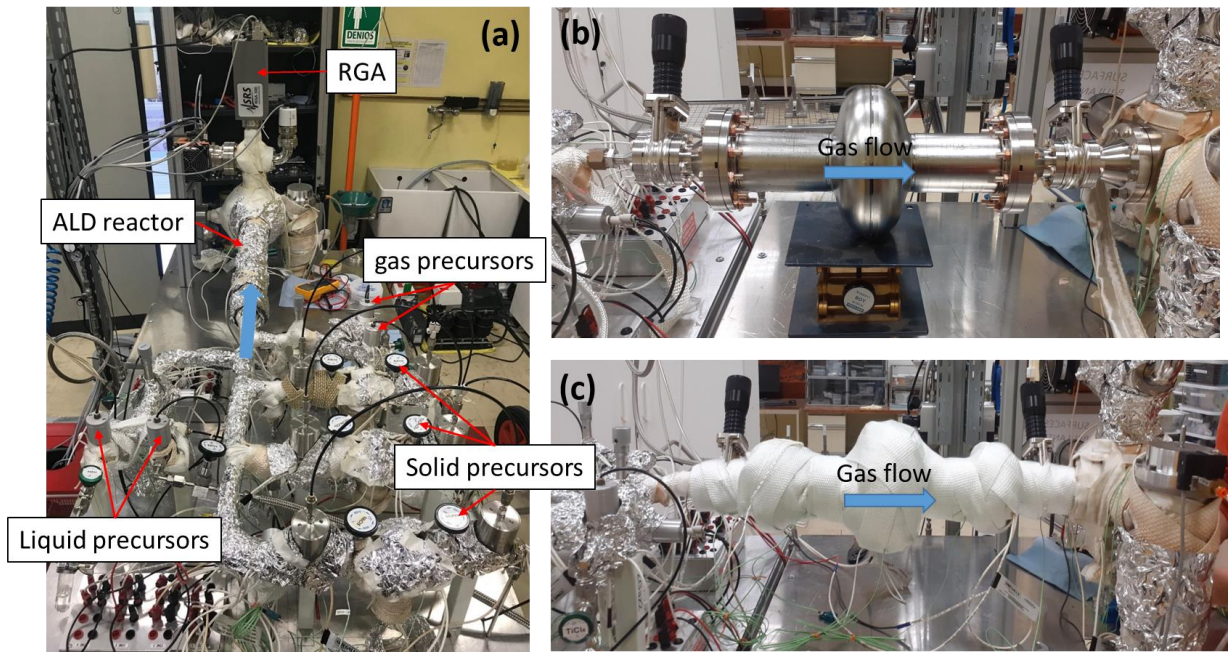

*Figure 1: (a) top view of the ALD system. (b) the 1.3 GHz cavity is installed on the ALD system, replacing the tube like ALD reactor chamber. (c) the cavity is dressed with thermocouples and heating resistance to control its temperature uniformity during the ALD growth. The blue arrows indicate the gas flow direction.*

The deposition parameters used for the Al$_2$O$_3$ (M. D. Groner, 2004) and TiN (L. Hiltunen, 1988) coatings are summarized in Table 1. The films were deposited on 500 µm thick Si wafer pieces. After the growth all the TiN/ Al$_2$O$_3$ films were cooled down in situ to 30°C prior to air exposure.

*Table 1: ALD growth parameters used.*

| Parameters | Al$_2$O$_3$ | TiN |
| --- | --- | --- |
| Growth temperature | 250°C | 450°C |
| Precursors | TMA + H$_2$O | TiCl$_4$ + NH$_3$ |
| Pulse/purges (s) | 1/15 + 1/15 | 2.5/15 + 0.5/15 |
| Number of cycles | 100 | Between 1 and 500 |
| Growth rate (Å/cycle) | 0.8±0.02 | 0.3±0.05 |

The initial 100 cycles of Al$_2$O$_3$ were deposited in order to insure a reproducible growth rate for the TiN films that can be very surface-dependent in particular in the nucleation regime of a few cycles.

The ALD on a 1.3 GHz niobium superconducting cavity was carried out on the same ALD reactor, replacing the ALD deposition chamber by the cavity itself (Figure 1), insuring a deposition only on the inside of the cavity. Prior to doing the deposition in a Nb cavity, the homogeneity of the film thickness along the cavity profile was measured by X-ray reflectivity (XRR) on Si samples placed along a test 1.3 GHz tesla shape aluminum cavity as shown in Figure 2.

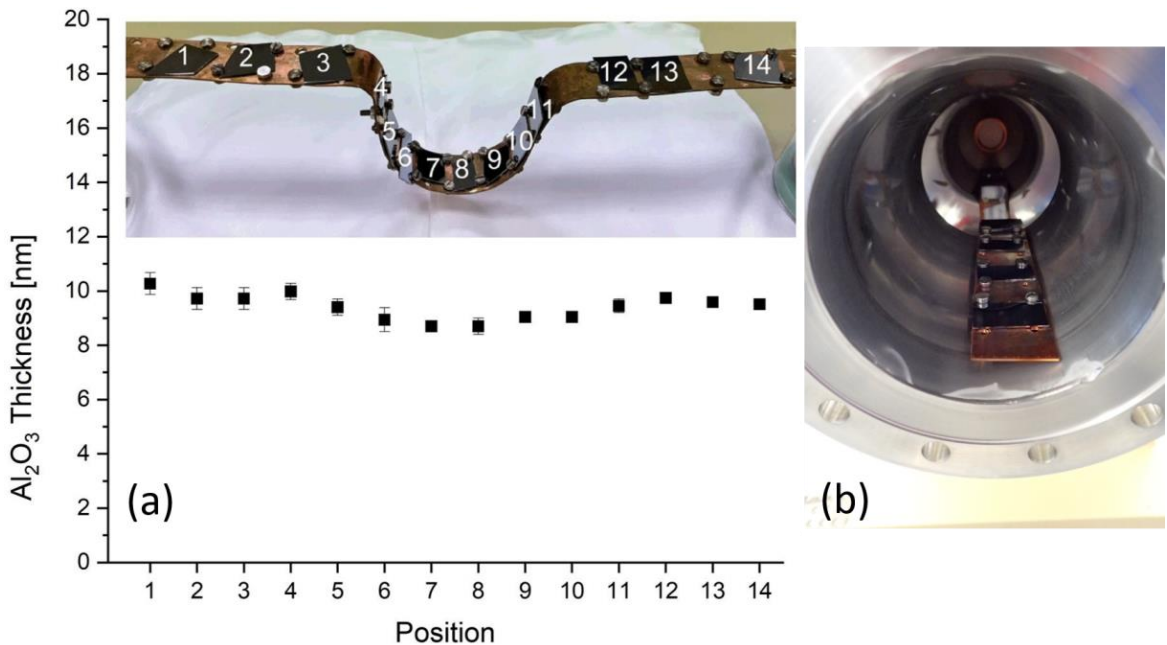

*Figure 2: (a) Thickness profile measured by XRR on Si samples placed on a custom sample holder placed inside a test Al cavity (b).*

The ALD deposition test consisted in 100 cycles of Al$_2$O$_3$ following the growth parameter listed in Table 1. The thickness profile homogeneity is very good with an average thickness, $d = 3 \pm 0.35$ nm and a growth rate per cycle (GPC) is agreement with literature data (M. D. Groner, 2004).

**2.2 Electron yield measurements:**

The TEEY measurements were performed in UHV facility (typically 1E-10 mbar), specially designed for SEY characterizations. The analysis chamber is equipped with Kimball Physics ELG-2, Omicron Mg k α/Al k α X-ray source, SIGMA hemispherical electron energy analyzer, Focus –FDG150 ion source and Kimball Physics low energy electron gun ELG2 (1 eV-2 keV).

The TEEY were measured using the protocol described in the reference (T. Gineste, 2015). In order to avoid additional conditioning of the surface during the TEEY measurement shorts electron beam pulses of 6 ms and a current of hundred nA were used. Each measurement is repeated and averaged 10 times per point (incident energy) with a typical variation of less than 2%. The incident electron beam was set normal to the surface. The surface conditioning was performed at 500 eV using a high current flood gun Kimball Physics FRA 2X1-2 electron.

**2.3 X-ray photoelectron spectroscopy measurement:**

The XPS experiments were performed in the same chamber used for the electron yield measurements described previously. The X-Ray source is at an incident angle of 60° with respect the sample normal, the emission current was 20 mA and the working distance between the analyzer and the sample was 30 mm. The analyzer acceptance angle was 20° and the pass energy filter was set to 20 eV for high resolution scans. These parameters were kept constant for all XPS measurements described in the paper.
The XPS spectra of the Ti 2p and N 1s core-level regions have been analyzed by the peak fitting software CasaXPS (N. Fairley, 2021), using a Shirley-type background and mixed Gaussian/Lorentzian peak shapes. The TiN component of Ti 2p region was fitted with a line shape extracted from the Ti 2p spectrum of a clean TiN surface deposited by ALD (D. Jaeger, 2013).

**2.4    X-ray reflectivity**

The films thicknesses, roughness and density were measured by X-ray reflectivity (XRR) using two 5-circle Rigaku Smartlab diffractometers with Cu Kα radiation coming from a rotating anod and two different optical setups. The incident beam was parallelized by a parabolic mirror, reducing the angular divergence to approximately 0.01°. Further reduction of the beam divergence was achieved either at the reception part by a parallel slit analyzer (PSA = 0.228°) or at the incidence part by a channel-cut 2-reflection Ge(220) monochromator which selected also Cu Kα1 radiation. 2.5° or 5° Soller slits parallel to the incidence plane were used to limit the transverse divergence of the beam and reduce scattering noise. The reflected beam was counted by a HyPix3000 detector or a NaI scintillation one. The XRR curves were then fitted using X'Pert Reflectivity software to extract the density, thickness and rugosity of the different layers.

**2.5    Simulations**

The secondary emission yield of the electrons has been computed between some eV up to some keV using a Monte Carlo simulation toolkit based on the MICROELEC module of the GEANT4 package (J. Pierron, 2017) (Q. Gibaru, 2021) (C. Inguimbert, 2022). This module has diverse applications, but to address the SEY modelling some specific physical processes have been accounted for. The quantum reflection/refraction interaction process has been integrated at vacuum/matter or matter/matter interfaces. This code also account for the Work Function (WF) of the material, which influences the behavior of electrons crossing material interfaces. Additionally, at low energies, it's crucial to consider the potential energy of electrons in the material, which becomes significant relative to the incident energy. This involves factoring in the binding energy of each atomic shell and the potential energy of weakly bound electrons in the valence and/or conduction bands. Interaction cross sections are evaluated for each inner shell (K, L, etc.), with a random selection used to determine interactions at each step. Weakly bound electrons in conduction and valence bands are treated similarly, assuming that the excited plasmon systematically dampens by emitting secondary electrons from the band. In this scenario, electrons are presumed to originate from an average energy level representing the band's width. Below some keV, the Monte Carlo code derives interaction cross sections from Optical Energy Loss Functions (OELF). Mermin's dielectric function provides a description of the momentum transfer. The elastic coulombian interaction cross-sections are determined using the ELSEPA code with the partial-wave method. For a deeper understanding of the code, readers can refer to the provided references (J. Pierron, 2017) (Q. Gibaru, 2021) (C. Inguimbert, 2022).

## 3. Results and discussion

### 3.1 Film Thickness and growth :

The thickness, density and roughness of the TiN/$Al_2O_3$ films were obtained using X-ray reflectivity curves and plotted as a function of the number of ALD TiN cycles noted N in Figure 3. For all the samples the thickness, density and roughness of the $Al_2O_3$ films was 85 ±0.5 Å/cycle, 3.2 ±0.2 g/cm$^3$ and 4 ±1 Å respectively.

We notice two growth regimes depending on the TiN film thicknesses:

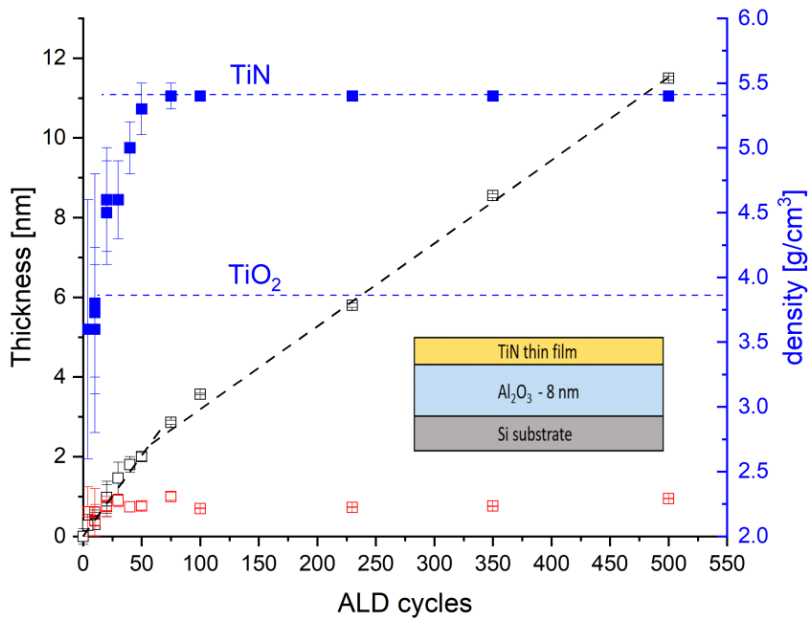

*Figure 3: TiN films thickness (black), roughness (red) and density (bleu) extracted from X-ray reflectivity measurements as a function of the number of ALD cycles of TiN.*

For N > 50: the film density remains constant and equal to 5.4 g/cm$^3$ which is consistent with the density of bulk-like TiN film found in literature (A. Satta, 2000). We observe a linear dependence of the thickness with number of ALD cycles with a growth rate per cycle (GPC) equal to 0.2 ± 5.10$^{-3}$ Å/cycle also in agreement with literature values (L. Hiltunen, 1988). The films roughnesses are constant around 7-9 Å.

For N ≤ 50, as the number of cycle decreases we notice that the density of the films gradually decreases towards the density of $TiO_2$ that ranges between 3,78 to 4.23 g/cm$^3$ depending on the crystalline phase (N.D. Johari, 2019). This behavior indicates the presence of titanium oxide in the film chemical composition. The growth rate is estimated to be 0.38 ± 0.05 Å/cycles, about twice higher than for the thicker films, which suggests an increased $TiCl_4$ reactivity on the hydroxyl terminated $Al_2O_3$ surface as compared to a fully covered TiN surface that occurs around 50 cycles or 2 nm. In this nucleation regime, the TiN film roughness is on the same order as the film thickness.

The secondary electron yield is extremely sensitive the outermost surface layers properties. It is therefore necessary to improve on the TiN film thickness measurements errors obtained by XRR for very thin films ≤ 50 cycles. X-ray Photoemission spectroscopy (XPS) was used to not only perform chemical composition analysis of the surface but also to estimate the TiN overlayer average thickness for these very thin films.

The Ti-2p and N-1s XPS spectrum are represented in Figure 4 (a) and (b) for various TiN film thicknesses from 0 to 100 cycles. The same Shirley background is used to extract the area of Ti 2p emission lines for each spectra that is then normalized by the areas obtained for the 50 Cy TiN film. We can see in Figure 4 (c) that the normalized areas of the core-level Ti-2p and N-1s increase continuously with increasing TiN ALD cycles, until it reaches a saturating value above 50 cycles. This saturation is caused by the finite escape depth of photo-electrons that ranges from a few nanometers for metals to about 10 nm for insulating materials (J. Zemek, 1995). As a results, beyond this threshold the film thickness continues to increase as shown in figure 2 whereas the XPS signal intensity saturates.

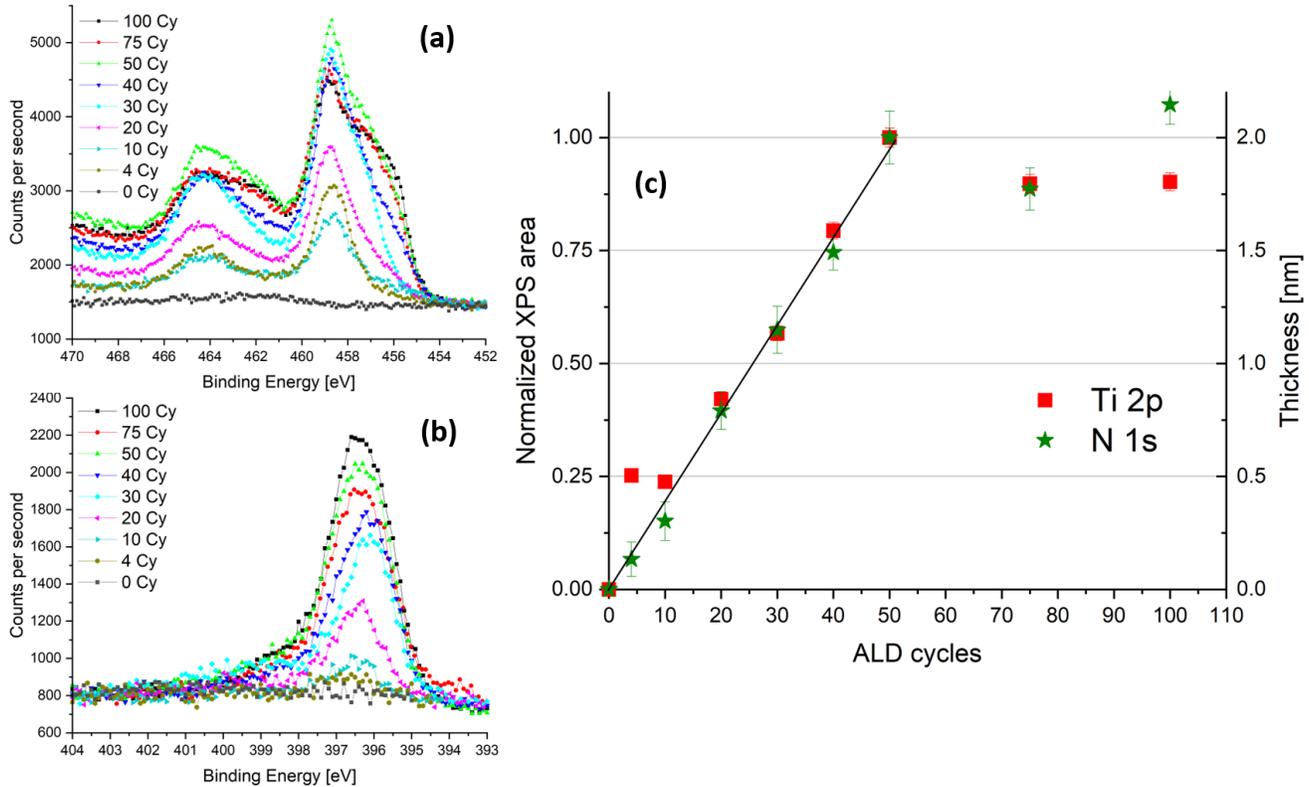

*Figure 4: (a) and (b) represents the Ti-2p and N-1s XPS spectrum measured for different TiN film thicknesses. (c) The area of each spectrum is normalized to the one obtained for 50 cycles of TiN.*

Based on XRR thickness values obtained for the 50 cycles TiN film and for which the fits gives reasonable errors (typically less than 5% of the TiN film thickness) we can calculate the average thickness of the films < 50 cycles with the simple relation: $d_{N\ cycles} = \frac{A_{N\ cycles}}{A_{50\ cycles}} \times d_{50\ cycles}$ where $A_{50\ cycles}$ and $A_{N\ cycles}$ are the area for a 50 cycles and N cycles thick TiN film, $d_{N\ cycles}$ and $d_{50\ cycles}$ are film thicknesses of the N cycles and the 50 cycles TiN films which is $2 \pm 0.1$ nm. The calculated growth rate up to 50 cycles obtained by this method is $0.40 \pm 0.02$ Å/cycles and the TiN film average thicknesses errors are at least a factor of 2 better than the XRR fitting analysis.

The results shown are for as grown films, prior to electron irradiation conditioning. The same procedure was applied after in-situ conditioning of 400 mC/mm$^2$ (not shown). We do not expect the total amount of Ti to be affected by the conditioning and it was assumed that the total thickness for 50 cycles remains at $2 \pm 0.1$ nm. Both set of data show very consistent thicknesses before and after conditioning for N ≤ 50 cycles. For the remaining of this work, the TiN film thicknesses are taken from the XPS analysis for films thinner than 50 cy (or 2 nm) and from the XRR measurements for high number of ALD cycles.

In order to understand the film growth regimes described previously, we propose a simple ALD growth model that calculate the GPC as a function of the number of ALD cycles based on two reaction probabilities of TiCl$_4$ molecules on 1/ a O-H terminated Al$_2$O$_3$ surface, named θ, and 2/ on a surface that reacted with the previous ALD cycle, called θ'. We assume a random distribution of nucleation sites and that NH3 pulse does not affect the reaction probabilities mentioned previously. Under those assumptions, we can demonstrate that the TiN growth rate, GR$_{TiN}$, can be written as a function of ALD cycle n with the analytical formula:

$$GR_{TiN}(\theta, \theta', \rho, n) = \rho\theta(1-\theta)^{n-1} + \rho\theta'(1-(1-\theta)^{n-1}) \qquad 1$$

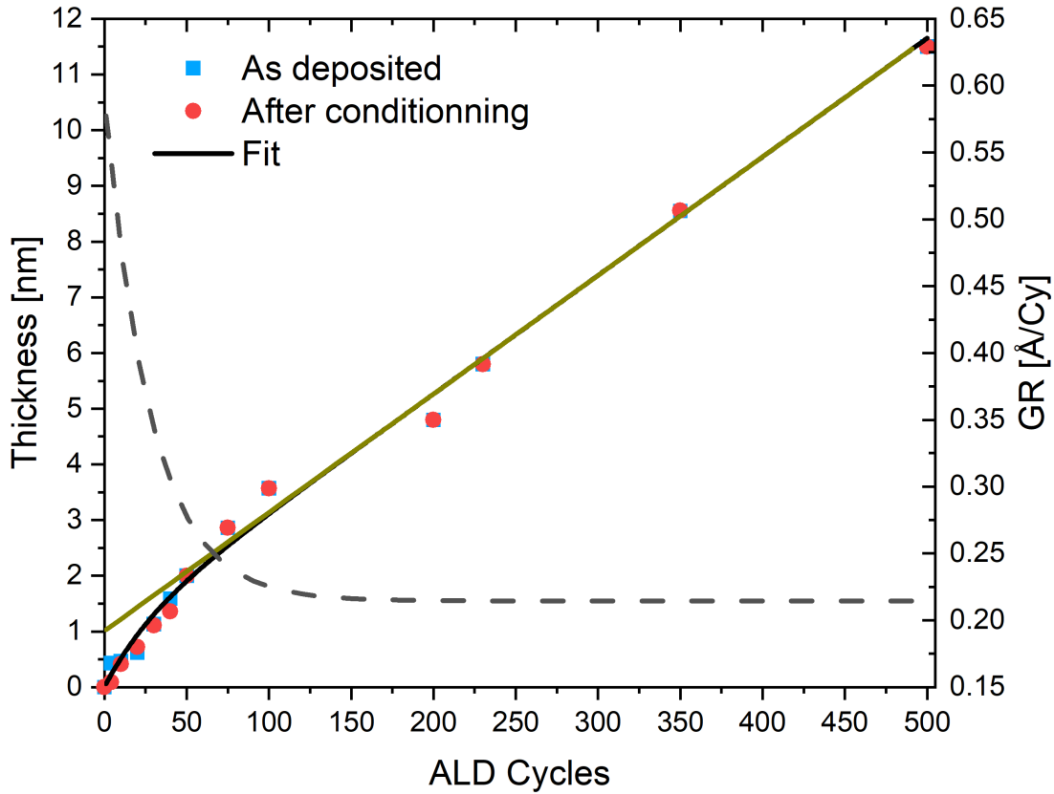

*Figure 5: Measured thickness with XPS and XRR as a function of ALD cycles for as deposited and conditioned TiN films. The dashed and solid lines are the fits using equation 1 for the growth rate GR (right axis) and equation 2 for the thickness (left axis).*

Where ρ is a quantum of average thickness. The left term of the sum describes the growth rate (GR) of the O-H terminated pristine surface and the right term the GR on a surface that already reacted with the previous ALD cycle. From equation 1 we can calculate the thickness, d, as a function of ALD cycles through the relation:

$$d = \int_0^N GR_{TiN}(\theta, \theta', \rho, n) dn$$

$$d = \rho\theta'N + \frac{\rho(\theta-\theta')}{(1-\theta)Ln(1-\theta)}((1-\theta)^N - 1) \qquad 2$$

Where N is the total number of ALD cycles. The fit using equation 2 shown in Figure 5 (black curve) gives the values ρ=16±0.5 Å, θ=0.035±1.10$^{-3}$ and θ'=0.013±5.10$^{-4}$. The growth rate can then be calculated as a function of the number of ALD cycles and is represented by the dashed curve in Figure 5 (right axis). In the limit $N \rightarrow \infty$, formula 1 reduces to the linear dependence represented by the dark yellow line:

$$d_{N\rightarrow\infty} = \rho\theta'N - \frac{\rho(\theta-\theta')}{(1-\theta)Ln(1-\theta)} \qquad 3$$

### 3.2 The total electron emission yield measurements :

The TEEY of the as deposited bilayer Al$_2$O$_3$-TiN was measured for various TiN films thicknesses. Figure 6 (a) and (b) show the TEEY measured as a function of primary electron energy from 0 and 1800 eV. All the TEEY curves shows a typical shape where the TEEY first increase with the energy of the primary electrons until it reaches a maximum value of TEEY $_{MAX}$ at energy E$_{max}$ and decreases again upon increasing the primary electron energy. Using the measured film thicknesses described in the previous section, we can extract the maximum TEEY as a function of TiN film thickness represented in Figure 6 c).

Starting from the bare 8.5 nm Al$_2$O$_3$ film, the maximum TEEY (TEE $_{MAX}$) is 4.65 ± 0.05 at around E$_{max}$ = 320 eV, in agreement with previous literature results with ALD Al$_2$O$_3$ (J. Guo, 2019). As the TiN thickness is increased, the TEEY$_{MAX}$ decreases exponentially to saturating values of 1.87 and 1.25 before and after conditioning for films thicker than ~ 2 nm or 50 ALD cycles (Figure 7).

Based on MICROELEC module, a code has been developed to simulate electron transport within a geometry consisting of an 8 nm Al$_2$O$_3$ substrate covered by a thin layer of TiN. The thickness of the TiN layer has been varied from 0.2 nm up to 9 nm. Contamination can also be accounted for thanks to an additional very thin layer of graphite. The thickness of this layer is defined to be 0.25 nm for the TiN films and 0.1 nm for the bare $Al_2O_3$ surface. The contaminated surface represents the conditioned sample while the as received sample is supposed to be graphite free. The simulations are summarized in Figure 6 c) and d). The volume is irradiated with normal incident electrons ranging in energy from 25 eV to 3 keV. All electrons produced during simulation and re-emitted by the surface are counted using a spherical detector surrounding the irradiated volume. Primary and secondary electrons can be distinguished to evaluate secondary emission and elastically backscattered yields. The impact of a change in the TiN layer thickness has been studied and compared to experimental data. TEEY vs. incident electron energy has been modeled for $Al_2O_3$ substrate topped with TiN thin films of various thicknesses.

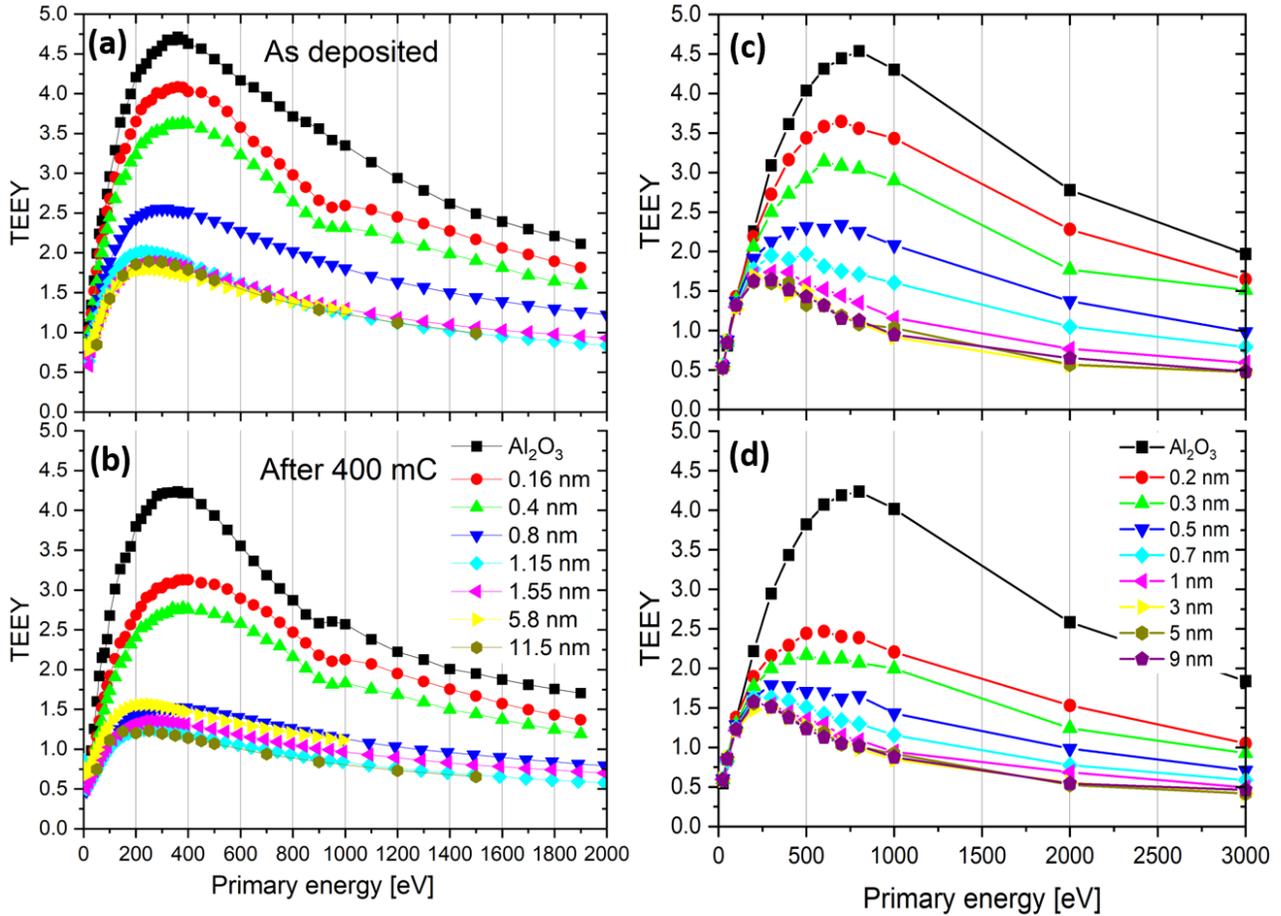

*Figure 6: experimental data of the TEEY curves before (a) and after (b) 400 mC/mm$^2$ conditioning for different TiN film thicknesses. (c) and (d) the corresponding numerical simulations using the model described in the text.*

The model reproduce well the experimental trends observed; an abrupt decrease of the TEEY max value (Figure 7) and a gradual shift of $TEEY_{MAX}$ energy towards lower primary electron energies as a function of the TiN film thickness (Figure 6 (c) and (d)). However, it is important to remain cautious with simulations and moderate our conclusions. Statistically, the Monte Carlo method is quite relevant for evaluating energy transfers related to each type of material. On the other hand, when material layers become extremely thin, they lose their uniformity, while the simulation assumes a consistently homogeneous layer. Even though the calculation remains relevant in terms of the probability of interaction with atoms of a given type, we can no longer assert that we have a good understanding of the work function of the material present on the surface, nor that we perfectly know the typical excitation energies, such as plasmons, which will be strongly affected by the structure of the surface layer. Nevertheless, considering all these precautions, we find that the simulation is in good agreement with the measurements.

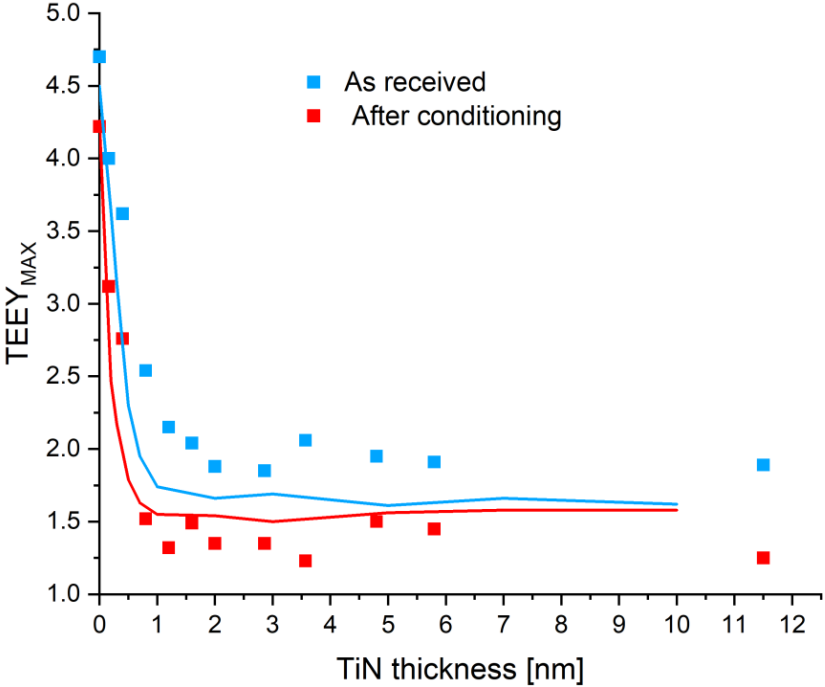

*Figure 7: Maximum value of the TEEY (TEEY$_{MAX}$) as a function of the TiN film thickness before and after conditioning (points) and the values extracted from the fits displayed in figure 6 (lines).*

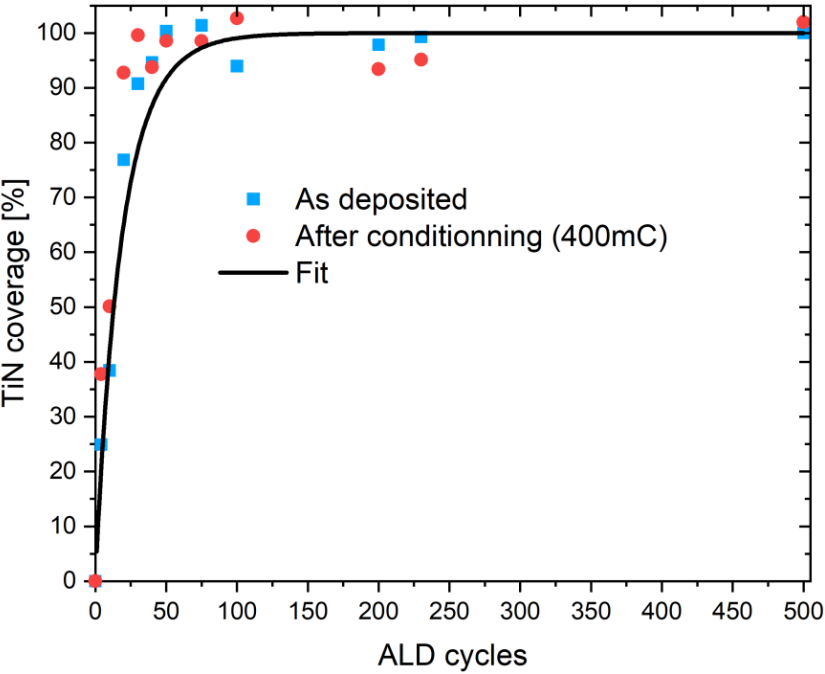

*Figure 8: TiN surface coverage calculated with equation 4 before (blue) and after (red) conditioning. The fit (black line) is calculated from the ratio of equations 2 and 3.*

The saturation of the $TEEY_{MAX}$ occurs at ~ 50 ALD cycles (or 2 nm) that corresponds to the change in the TiN growth rate shown in Figure 5. This observation further supports the nucleation scenario of TiN film on $Al_2O_3$ mentioned previously; the partial coverage of the $Al_2O_3$ surface by TiN up to 50 ALD cycles (or 2 nm) suggests that the $TEEY_{MAX}$ value measured is a combination of $Al_2O_3$ $TEEY_{MAX}$ and TiN $TEEY_{MAX}$ from 0 to 50 cy. From the dependence of $TEEY_{MAX}$ as a function of the ALD average TiN film thickness and the $TEEY_{MAX}$ values measured on bare $Al_2O_3$ and bulk TiN we can extract the TiN surface coverage as a function of number of TiN ALD cycles N using the formula:

$$TiN\ coverage\ [\%] = \frac{(1-TEEY_{MAX}(N)/TEEY_{MAX}(N=0))}{(1-TEEY_{MAX}(N\rightarrow\infty)/TEEY_{MAX}(N=0))} \qquad 4$$

The dependence of the TiN surface coverage as a function of the TiN average film thickness for the as deposited and after conditioning are represented in Figure 8. The TiN coverage can also be calculated from the ratio: $d/d_{N\rightarrow\infty}$ given by equations 2 and 3 and represented by the black solid line in Figure 8. The agreement with the data is surprisingly good considering the simplicity of the nucleation model proposed.

### 3.3 Chemical composition analysis and film resistivities.

In addition to the TiN film thickness, the TEEY is also sensitive to the surface chemical composition. In order to investigate the thin films chemical composition and its thickness dependence, XPS measurements were performed in-situ prior to the TEEY measurements before and after 400 mC *in-situ* conditioning.

The Ti chemical composition was extracted from the fits of the Ti 2p peaks regions between 454 and 468 eV with the same fitting procedures described in the supplementary information; three mains Ti peaks oxidation states were necessary to correctly fit the data: $TiO_2$, TiNO and TiN. The results from the fits of these three oxidation states as a function of the TiN film thickness are displayed in figure 8 (a). The chemical composition of TiN films shows significant dependence on the layer thickness: for very thin films ≤ 0.5 nm the Ti is essentially in the form of $TiO_2$, between 0.75 to 1.7 nm the film is composed of $TiO_2$ and TiNO with negligible amount of TiN. As the film thickness increases above 2 nm, all oxidation states tends towards 30%. After *in-situ* conditioning with 400 mC (open symbols in figure 8) the TiN component of the Ti 2p peak region increases whereas the $TiO_2$ decreases for films thicknesses above 0.5 nm. This trend can be interpreted as the partial reduction of the $TiO_2$ upon electron surface irradiation as it was observed previously in a number of oxide (O. Dulub, 2007) (J.-W. Park, 1996).

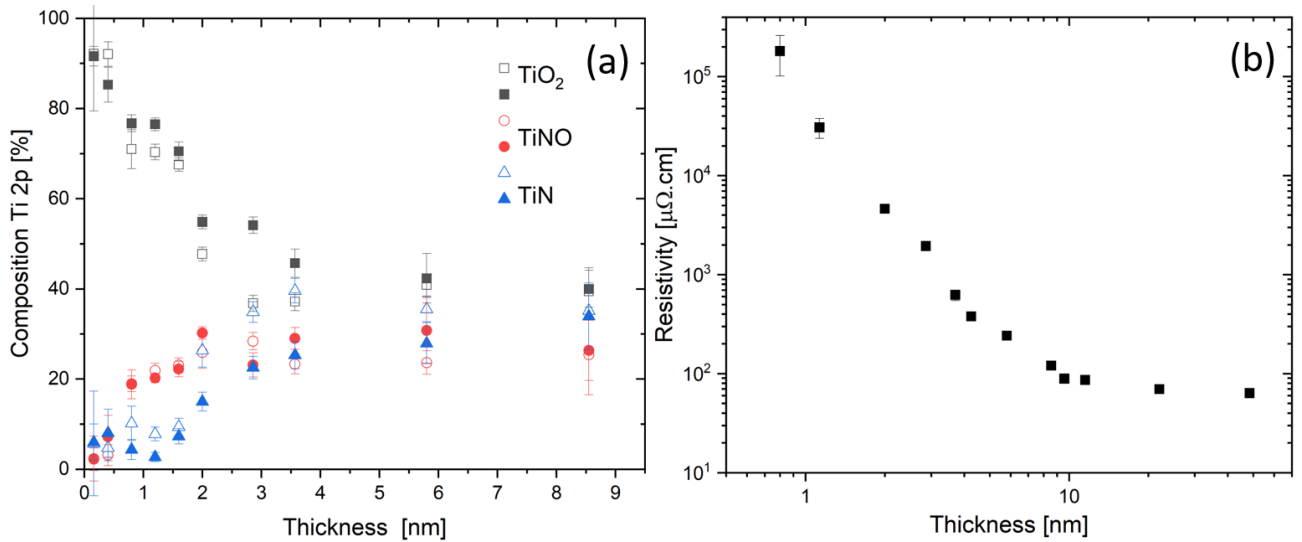

*Figure 9: Composition of the Ti 2p spectrum as a function of the TiN film thickness. Close and open symbols corresponds to measurements before and after 400 mC/mm² in-situ conditioning.*

The resistivity of the as-deposited TiN films displayed in figure 8 (b) were measured with four point probes at room temperature. The bulk resistivity is 63 µΩ.cm and start increasing below 10 nm. For films thinner than ~ 0.8 nm the resistivity values were too higher for our set up to be measured (> $10^6$ µΩ.cm). Attempts to fit the data with Fuchs (Sondheimer, 1952) or Mayadas (A.F. Mayadas, 1970) theories that takes into account various electron diffusion mechanisms (grain boundaries, point defect and surfaces scattering) did not succeed because the experimental resistivity values increase faster than what is predicted by the theory. This can be explained by the change in chemical composition probed by XPS with an increasing insulating $TiO_2$ component as the film thickness decreases.

### 3.4 Radio frequency tests on 1.3 GHz superconducting cavities.

In order to test the multipacting mitigation approach using a thin TiN layer deposited by ALD, a first attempt was made to deposit 10 nm of $Al_2O_3$ inside a 1.3 GHz cavity followed by a post annealing at 650°C for 10 hrs in high vacuum. The purpose of such deposition and annealing procedure is described in (Y. Kalboussi, 2024). This process was repeated twice on the same cavity with a reset on the surface by chemical etching (J Knobloch, 1997) in between the $Al_2O_3$ ALD depositions and post annealing. For both RF tests after deposition and post annealing, a strong multipacting barrier reproducibly occurred at 15-18 MV/m and could not be processed, preventing us from reaching higher accelerating fields.

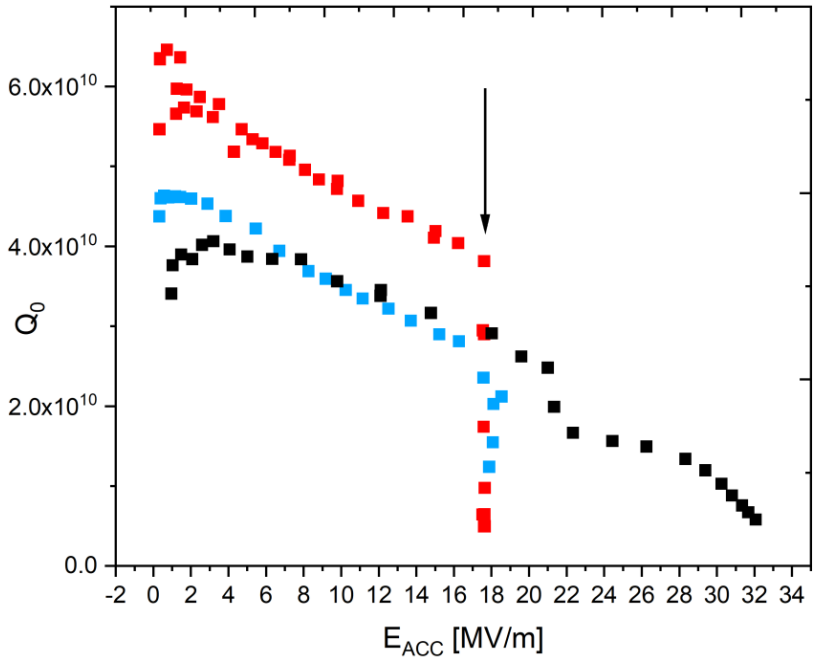

*Figure 10: RF tests of a 1.3 GHz Nb cavity baseline (black points) with two 10 nm $Al_2O_3$ coatings and post annealing (red and blue points). The Multipacting barrier is indicated by the black arrow.*

The baselines RF test measured prior to the ALD deposition (black curve in figure 9) reached a maximum $E_{ACC}$ of 32 MV/m with a moderate multipacting barrier present around 21 MV/m and easily processed as expected for Nb cavities (Padamsee, 2017). In order to determinate the cause for this strong multipacting barrier, numerical simulation were carried out using Superfish (Halbach) and Fishpact (E. Donoghue) codes to calculate the electromagnetic field distribution and the electron trajectories in a single cell elliptical cavity. The electron energy was calculated at each impact with the surface, and the TEEY curves measured (see Figure 11 (a)) for different surface chemical composition (bare Nb, $Al_2O_3$/Nb and TiN/$Al_2O_3$/Nb) on cavity grade Nb coupons were used to calculate the number of secondary electrons emitted as a function of the $E_{ACC}$ in the 1.3 GHz geometry. For more details see ref (Kalboussi, 2023).

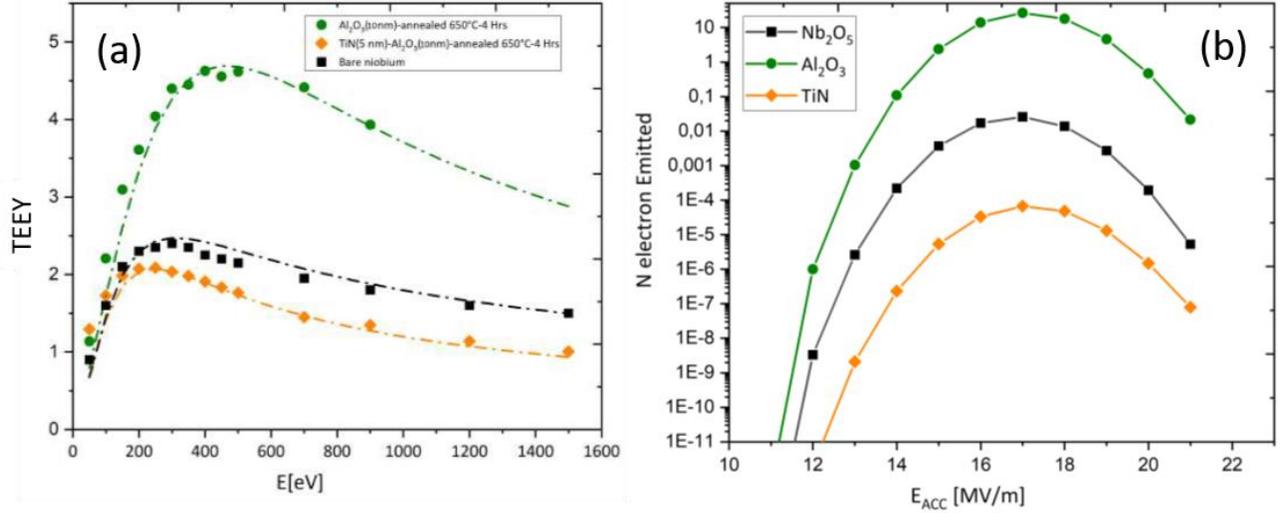

*Figure 11: (a) TEEY curves measured as a function of incident electron energy for the various surface composition indicated. (b) The corresponding numerical simulations of the number of electron emitted into the SRF cavity as a function of accelerating field.*

The numerical simulations (Figure 11 (b)) reveal a maximum of emitted secondary electrons between 16 and 19 MV/m in very good agreement with the experimental data (Figure 9). In addition, the simulation indicates three orders of magnitude higher electron emission with a $TEEY_{MAX}$ of $Al_2O_3$ of 4.6 as compared to bare Nb surface with a $TEEY_{MAX}$ of 2.3.

This indicates that the large $TEEY_{MAX}$ value of the $Al_2O_3$ coating is responsible for the observed multipacting barrier. Moreover, it suggests that coating the cavity inner-walls with a 5 nm TiN film deposited on top of a 10 nm thick $Al_2O_3$ film can reduce the multipacting probability by 6 orders of magnitude compared to that in the $Al_2O_3$-coated cavity and therefore should suppress the multipacting effect encountered. The number of emitted electron can even be reduced compared to a bare niobium cavity.

In order to put to the test the numerical simulations, we deposited a 5 nm TiN film (230 ALD cycles) inside the previously tested $Al_2O_3$-coated cavity (blue curve in figure 9 and Figure 12) and tested its performance under RF fields as shown in Figure 12. The RF performances (green curve) deteriorated significantly with respect to the $Al_2O$-coated cavity baseline; the quality factor is decreased by over 2 orders of magnitude down to $10^8$ and correspondingly the surface resistance increased 330 times to 2200 nΩ. The measurement errors on the Q for such poor values are important (~ 100%) due the large mismatch between the power transmission and the cavity.

The presence of a thin, non-superconducting, metallic layer within the RF penetration depth can affect drastically the dissipation. One can calculate this effect with the relations:

$$P_{surf} = \int_0^\infty \langle P_{vol} \rangle dz = \int_0^d \langle P_{TiN} \rangle_{Normal}\, dz + \int_d^\infty \langle P_{Nb} \rangle_{Supra}\, dz$$

$$= \frac{H_0^2}{2}\left(\frac{1-e^{-d/\delta_{TiN}}}{\sigma_{TiN}\delta_{TiN}} + Re\left[\frac{1}{\sigma_{nb}\lambda_{Nb}}\right]\right) \approx \frac{H_0^2}{2} R_{S\,TiN}$$

Where $P_{surf}$ is the RF power dissipated in the internal surface of the cavity, $\sigma_{TiN}, \sigma_{Nb}, \delta_{TiN}, \lambda_{Nb}$ are the TiN and Nb conductivity and penetration depths normal and superconducting respectively. d is the TiN film thickness and $H_0$ is the RF magnetic field amplitude at the surface.

The resistivity value (figure 8(b)) of a 5 nm TiN film is ~ 300 µΩ.cm and we can calculate $\delta_{TiN}$ ~ 20 µm at 1.3 GHz and therefore d << $\delta_{TiN}$. In this limit, we have:

$$R_{S\,TiN} = \frac{d}{\sigma_{TiN}\delta_{TiN}^2}, \quad \delta_{TiN} = \sqrt{\frac{2}{\omega\,\mu_0\,\sigma_{TiN}}}$$

$$R_s = R_{S\,TiN} = \frac{d\,\omega\,\mu_0}{2} \qquad 5$$

Hence in the very thin film limit the surface resistance $R_S$ does not depends on the intrinsic material layer properties such as the conductivity and is directly proportional to the film thickness. The numerical estimate gives $R_S \sim 10^{-5}\,\Omega$ and the $Q = \frac{G}{R_S} = \frac{271}{10^{-5}} \sim 3 \times 10^7$ which are values consistent with the RF tests measurements. The presence of a metallic film on top of a high quality superconducting material (Nb) drastically reduces the resonator performances.

In order to diminish the Ohmic losses induced by the TiN film, we selected an optimal thickness that would present both a high resistivity and low TEEY$_{MAX}$ values i.e. below or equal to the bare Nb surface ~2.3 (black curve in Figure 11 (a)). For the second coating the TiN film thickness was decreased down to

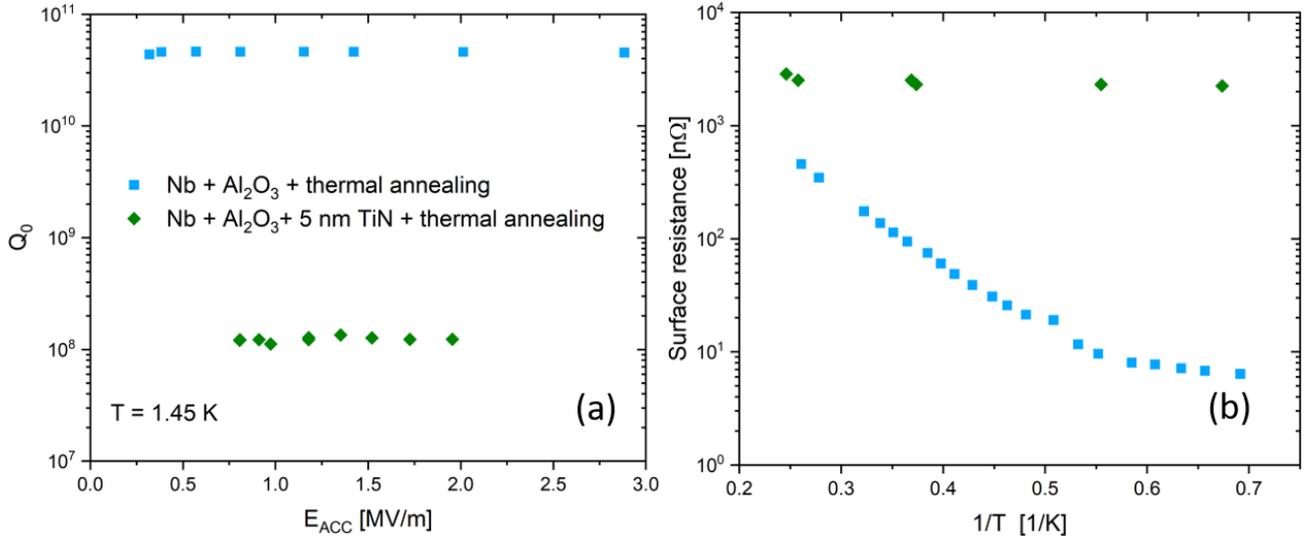

*Figure 12: RF tests at 1.45 K of a bare Nb cavity coated with 10 nm Al2O3 + 650°C post annealing (blue) and then with a 5 nm of TiN film (green). (a) Q vs E$_{ACC}$ and (b) the corresponding surface impedance vs temperature.*

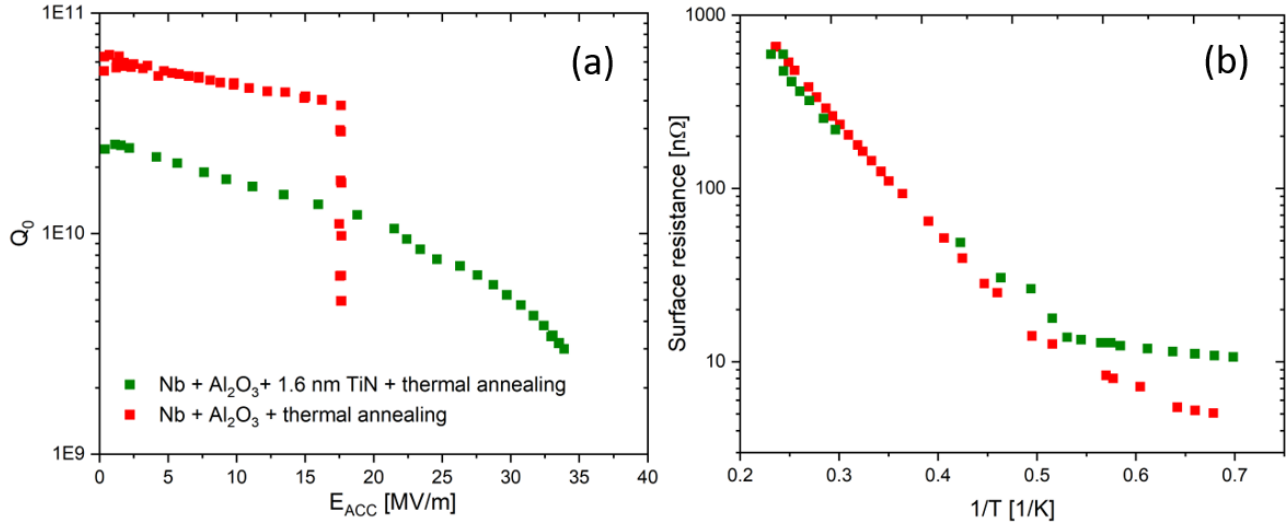

*Figure 13: RF tests at 1.45 K of a coated Nb cavity with 10 nm of $Al_2O_3$ and post-annealed at 650°C for 10 hrs in high vacuum (red) and then with 1.6 nm of TiN (green).(a) Q vs $E_{ACC}$ and (b) surface impedance vs temperature.*

40 cy ~ 1.6 nm that corresponds to the minimal thickness before which the $TEEY_{MAX}$ increase above 2. The film resistivity for this thickness increases by about two orders of magnitude (figure 8 (b)) to ~ $2.10^4$ $\mu\Omega$.cm as compared to the 5 nm thick TiN film and can hardly be considered as metallic anymore. Hence the formula 5 does not hold and the more pronounced dielectric nature of the thin TiN film might lower significantly the surface dissipation in the medium to high RF field amplitude regime.

The second RF test shown in Figure 13 (a) reveal that the quality factor is now in the $10^{10}$ range, two orders of magnitude higher than for the previous RF with a 5 nm TiN coating. Correspondingly, the surface resistance decreases from 2200 n$\Omega$ down to 10.8 n$\Omega$ at low temperature (Figure 13 (b)) upon reducing the TiN thickness from 5 to 1.6 nm. Importantly, the multipacting barrier at 18 MV/m disappeared, extending the range of accelerating gradient from 18 MV/m up to 35 MV/m and recovering the bare Nb RF maximal $E_{ACC}$ performance. These results demonstrate the ability of ALD to successfully mitigate the multipacting in SRF cavities by tuning both the $TEEY_{MAX}$ and the film resistivity owing to the unique ALD film thickness control capability on complex shape objects.

## Conclusion

In conclusion, we have investigated the TiN film thickness dependence of the TEEY, chemical composition and resistivity. The ALD growth mechanisms explained by a simple model lead to an incomplete surface coverage of the TiN for films thinner than 1.5-2 nm. Consequently, the $TEEY_{MAX}$ values can be understood as a linear combination of the $Al_2O_3$ and the TiN $TEEY_{MAX}$ respective values. For film thicker than 2 nm, the surface is fully covered with TiN and the $TEEY_{MAX}$ values saturate. The TiN film resistivities increase exponentially below 10 nm ranging from 63 $\mu\Omega$.cm in the bulk limit to over $10^5$ $\mu\Omega$.cm for 1 nm thick film. The careful selection of a suitable set of $TEEY_{MAX}$ and resistivities values by finely tuning the TiN film thickness enabled the suppression of the multipacting phenomena observed in SRF cavities while maintaining the high quality factor of the superconducting resonators in the $10^{10}$ range. The scalability of ALD from coupons to real device, is an opportunity to apply this method to other particle accelerator devices such as drift tubes or power couplers, and other RF devices used in satellites for instance with their own optimal resistivities and TEEY values requirements.

## Authors declarations

**Conflict of interest:** The authors declare no conflict of interest.

**Authors contributions: Yasmine Kalboussi:** Conceptualization; Data curation; Formal analysis; Investigation; Methodology; Resources; Validation; Visualization; Writing–original draft. **Sarah Dadouch:** Investigation. **Baptiste Delatte:** Resources. **Frederic Miserques**: Investigation. **Diana Dragoe:** Investigation. **Fabien Eozenou:** Resources. **Matthieu Baudrier:** Resources. **Sandrine Tusseau-Nenez:** Investigation, Writing–original draft. **Yunlin Zheng:** Investigation; Writing–original draft. **Luc Maurice:** Resources. **Enrico Cenni:** Resources. **Quentin Bertrand:** Resources. **Patrick Sahuquet:** Resources. **Elise Fayette:** Resources. **Gregoire Jullien:** Resources. **Christophe Inguimbert:** Investigation, Software, Writing–original draft. **Mohamed Belhaj:** Investigation; Data curation; Writing–original draft. **Thomas Proslier:** Conceptualization; Data curation; Formal analysis; Funding Acquisition; Supervision; Investigation; Methodology; Validation; Visualization; Writing–original draft.

## Acknowledgements

The authors would like to thank Thierry Pepin-Donat from IJCLab for providing a HV thermal treatment and Claire Antoine from CEA for insightful discussions. This project has received funding from the region Ile de France project SESAME AXESRF, the European Union's Horizon 2020 Research and Innovation programme under Grant agreement No 101004730 and Grant agreement No 730871.